# Ecological patterns of genome size variation and the origin of species in salamanders

Bianca Sclavi [1*] and John Herrick [2**]

Author affiliations:
1. LBPA, UMR 8113, ENS de Cachan, Cachan, France
2. Department of Physics, Simon Fraser University, 8888 University Drive, Burnaby, British Columbia VSA 1S6, Canada

   * Corresponding author
   ** Corresponding author at 3, rue des Jeûneurs, 75002 Paris, France.

   E-mail addresses: jhenryherrick@yahoo.fr (J. Herrick), sclavi@lbpa.ens-cachan.fr (B. Sclavi).

*Abstract:* Salamanders (urodela) have among the largest vertebrate genomes, ranging in size from 10 to over 80 pg. The urodela are divided into ten extant families each with a characteristic range in genome size. Although changes in genome size often occur randomly and in the absence of selection pressure, non-random patterns of genome size variation are evident among specific vertebrate lineages. Here we report that genome size in salamander families varies inversely with species richness and other ecological factors: clades that began radiating earlier (older crown age) tend to have smaller genomes, higher levels of diversity and larger geographical ranges. These observations support the hypothesis that urodel families with larger genomes either have a lower propensity to diversify or are more vulnerable to extinction than families with smaller genomes.

*Introduction*

Genome size in vertebrates varies more than two hundred fold from 0,4 picograms (pg) in pufferfish to over 120 pg in lungfish (*1*). Evolution of genome size is believed to be due to the non-adaptive consequences of genetic drift with most if not all of the variation corresponding to differences in non-coding DNA such as transposable elements (*2-4*). Genome size therefore reflects the balance between the passive gain and loss of neutral or nearly neutral DNA sequences during the course of evolution (*5*). This balance is related to the effective population size, with genome size tending to increase inversely with the effective population size (*4, 6*).

In contrast, non-neutral changes in genome size are expected to coincide with speciation events. One recent proposal suggests that speciation events occur as a result of relatively rapid amplifications in genome size followed by long periods of slow DNA loss (*7*). According to this scenario, speciation corresponds primarily to the acquisition of new genes and new regulatory sequences. Expansions in genome size beyond a certain level, however, are considered maladaptive (*8*). Phylogenetic lineages in plants with large genomes, for example, correlate with low levels of species-richness (*9, 10*).

In amphibians, biodiversity generally depends on latitude: the highest biodiversity occurs in amphibian families located near the equator (*11*). The latitudinal gradient has been attributed to differences in speciation and extinction rates between the tropics and temperate zones (*12*). Lineages near the tropics have higher than expected rates of speciation and lower than expected rates of extinction, while the reverse was found for lineages occupying temperate zones (*12*). Rates of diversification therefore increase along the latitudinal gradient resulting in higher species richness towards the equator.

To date, some 655 salamander species have been identified and grouped into ten distinct families. Although salamander families tend to be poor in the number of species they contain, some salamander families such as the Plethodontidae have exceptionally high levels of diversity (*13*). The Bolitoglossinae and Plethodontinae lineages are the two most species rich genera in the Plethodontidae (*8*), while the Bolitoglossinae account for most of the species diversity among salamanders in the tropics (*13*). The latitudinal gradient in salamander species richness remains unexplained, but ecological factors such as environmental energy and carrying capacity are believed to play important roles in promoting species diversity (*14, 15*).

Life history traits such as body size and generation time have also been invoked to explain variations in species richness (*16, 17*). Although life history traits frequently correlate with species richness (*18*), the exact nature of the relationship remains unclear. Physiology, which is closely related to life history traits, is also an important factor in determining a species adaptive response to the environment. Salamanders, for example, have very low metabolic rates compared to other vertebrates (*13*).

One of the principal factors influencing life history traits in salamanders is their exceptionally large genomes compared to other vertebrates (*19, 20*). Cell size is tightly correlated with the amount of nuclear DNA in the genome (*21*), and exceptionally large cells have imposed constraints on salamander development and physiology (*20, 22*). Large cells in salamanders, for example, resulted in simplified brains (*23*). Large genomes, via their impact on cell size, can also have a pronounced effect on metabolic rate (*24*), which might explain the low metabolic rates found in salamanders.

Although variations in genome size primarily reflect neutral or non-adaptive mutational processes, genome size clearly has an effect on other traits that are likely subject to and influenced by natural selection. Developmental time in Plethodontidae, for example, is positively correlated with genome size (*25*). The influence of genome size on species richness and rates of speciation in vertebrates, however, remains an open question. In the following, we address the relationship between genome size, species richness and evolutionary age in salamanders, and examine how these and other parameters such as area and climatic niche rate interact in influencing the course of salamander evolution.

**Results**

*Species richness is negatively correlated with C-value.*

The origin of urodela dates from 155 to 170 Mya (*26*). Urodela inhabit a wide variety of ecological niches and exhibit a large diversity of life history traits, including small and large body sizes, paedomorphy, neoteny, metamorphosis and direct development (*13*). In an earlier study in amphibia, Pyron and Wiens revealed a number of ecological correlates between species richness and variables such as geographical latitude, environmental energy and climatic niche rate (*12*). Species richness in frogs, salamanders and caecilians also varies according to abiotic factors such as humidity and temperature (niche composition), and biotic factors such as productivity and rates of diversification (extinction and speciation). Figure 1 shows the phylogenetic tree from Pyron and Wiens that was used here to investigate the relationship between genome size in salamanders and these ecological factors (*27*).

We examined genome size variation in salamanders as a function of the different parameters from the Pyron and Wiens dataset and C-values from the Animal Genome Size Database (*1*). We found a strong correlation between genome size and species richness that appears to divide salamanders into two broad classes in a genome size dependent manner (adjusted $R^2$ = 0.58, p=0.006; Figure 2A). We controlled for phylogeny using independent contrasts (*28, 29*), and found that this division of salamander families into two distinct classes is due to genome size and not to phylogenetic relatedness (adjusted $R^2$ = 0.84, p=0.0003; Figure 2B). Importantly, standardized contrasts indicate evolutionary rates (*5*). Therefore the rate at which species richness changes correlates strongly with the rate at which genome size changes, suggesting that the evolution of species in salamanders is contingent upon changes in genome size (see discussion).

Indeed, large and correlated differences in both genome size and species richness in extant salamander families is apparent from the phylogenetic tree. Some sister taxa have widely different genome sizes and differing species richness despite being most closely related (Figure 1). The Hynobidae, for example, have the smallest average C-value while their sister taxon the Cryptobranchidae have much larger genomes and correspondingly lower species richness. Likewise, the Amphiumidae have among the largest salamander genomes and very low species richness, whereas their sister taxon the Plethodontidae have among the smallest average C-value and the highest species richness. The sister taxa Ambystomatidae and Dicamptodontidae exhibit a similar trend between genome size and species richness. We conclude that genome size is a factor that negatively impacts species richness in salamander families.

*Species richness is negatively correlated with body size*

Our observation of a negative correlation between species richness and C-value might be related to a number of different life history traits that genome size directly or indirectly influences, or conversely that influence genome size. Although several possibilities exist including effective population size, abundance and generation time, we focus here on one of the known major correlates of species richness in metazoa: body size (*30*). We consulted the AmphibiaWeb database for snout to vent lengths (SVL) for each of the ten families (*31*), but did not find a strong correlation between average C-value and SVL measurements (adjusted $R^2$ = 0.24, p=0.1; Figure 3A; Table1).

The correlation between independent contrasts of C-value and body size, however, revealed a stronger correlation between these two variables (adjusted $R^2$ = 55%, p=0.01; Figure 3B), suggesting that larger differences in C-values are more closely associated with larger differences in body size. Hence, while genome size and body size are weakly correlated, evolution of genome size and body size tend to be more positively associated: as genome sizes increase (or decrease) during species diversification, body size tends to increase (or decrease) as well.

Species richness, in general, is not consistently correlated with body size in all examined metazoa, for example mammals (*16*). We found, however, that in salamanders body size has an impact on species richness (adjusted $R^2$ = 0.44 p=0.02; Figure 3C). Controlling for phylogeny using independent contrasts revealed a relatively stronger correlation between body size and species diversity (adjusted $R^2$ = 0.52, p=0.02; Figure 3D). Multiple regression (adjusted $R^2$ = 0.67, p-value = 0.008) shows that C-value is a more significant predictor than body size (p-value = 0.03 and 0.1, respectively; Table 2). Partial correlation analysis revealed that the two predictor variables, C-value and body size, together explain 74 % of the variance ($R^2$ = 0.69) in species richness in urodela families (Table 3). C-value and body size, however, have different contributions to the variance in species richness (85% and 68% of $R^2$, respectively) with unique contributions accounting for 33% and 16% of $R^2$ (Table 3).

Table 1 nevertheless indicates that genome size is more strongly associated with species richness than is body size when phylogeny is taken into consideration (C-value: F = 43, p = 0.0003; body size: F = 9.61, p = 0.02). Based on these findings, we suggest that the impact of C-value on species richness is operating through life-history traits and other variables in conjunction with body size. We conclude that genome size and body size have independent effects on species richness in salamander families with C value being the main predictor variable in explaining variance in species richness.

*Ecological correlates with genome size: time*

Assuming a nearly constant rate of speciation, species richness is expected to increase over time (*32, 33*). Older families should therefore contain proportionally more species. Conversely, the older the family the more likely lineages will die out over time. Hence, extant species richness in a family reflects the balance between speciation and extinction events. Phylogenetic stem age corresponds to the time of origin of a given salamander family; and, consequently, older families have experienced longer periods of mutation and genome size evolution. Salamander genomes are exceptionally large among vertebrates because of a mutational insertion bias (*34*), suggesting that older families will passively accumulate more DNA than younger families. Consistent with other observations on stem age and species richness (*35*), we found no significant correlation between stem age and genome size (Figure 4A).

Crown age, in contrast, corresponds to the formation of a clade, and hence to the beginning of an adaptive radiation. The longer the clade has existed the more time has passed for new species to appear in the clade (*33*). Similarly, the more species a clade contains the greater the likelihood of new speciation events (Yule-Simon process). Older clades (crown age) should therefore contain proportionally more species. Figure 4B shows a strong negative correlation ($R^2$ = 0.70) between average genome size and the phylogenetic crown age in the different urodela families. Older clades are therefore associated with smaller genome sizes on average. This observation is consistent with the finding that families with smaller genomes tend to be more species rich, and thus suggests that families with smaller genomes might experience more extensive adaptive radiations.

An examination of Figure 4A supports this suggestion at least at the family level. The family Sirenidae have the oldest stem age at 199.5 Mya, and an average genome size that corresponds approximately to the median size of the ten families (50 pg). The Chryptobranchidae, which have a stem age of 164.5 Mya, also have an average genome size that is approximately the median value (50 pg; Figure 4A). The two distinct classes of salamander families evident in Figure 2A correspond to the obligate paedomorphs including the Sirenidae, Cryptobranchidae, Proteidae and Amphiumidae, which have genome sizes larger than 50 pg on average, whereas families comprising metamorphic, facultative paedomorphs and direct developing species tend to have average genome sizes that are less than 50 pg. We propose these two distinct classes represent different evolutionary trajectories at the family level that correspond to different modes of genome size evolution: expansion and contraction.

*Ecological correlates with genome size: area*

Figure 4C shows a strong relationship between C-value and geographical area: lineages with smaller genome sizes occupy larger geographical areas. The Hynobidae, with the smallest average genome size, inhabit the broadest geographical range with the exception of the Salamandridae, which span the largest area (over 14,000,000 $Km^2$). Similarly, the Ambystomatidae and the Plethodontidae, which have average C-values of about 30 pg, occupy extensive areas of significantly different sizes: 11,000,000 and 7,000,000 $Km^2$, respectively. Families with smaller genomes (< 50 pg) therefore exhibit a 2X difference in the respective areas they inhabit.

In contrast, families with genome sizes greater than or equal to 50 pg appear to be restricted to areas that have approximately the same range (2,000,000 $Km^2$). Among these families the Proteidae occupy the largest geographical area (3,000,000). This area is significantly smaller than the area occupied by the Plethodontidae (7,000,000), which is the smallest area occupied by salamanders with C-values < 50 pg. Hence, a 2X difference in area exists between families with larger genomes compared to families with smaller genomes.

These observations indicate that dispersal in salamanders with genome sizes >= 50 pg is severely restricted compared to families with smaller genomes. A restricted geographical area might explain the lower species richness in salamander families with larger genomes; or conversely, an intrinsically lower species richness might have imposed a constraint on the extent of a family's dispersal. Comparing the similarity of Figure 2A and Figure 4C, area appears to be the ecological correlate that best accounts for a family's species richness. These two parameters were previously found to correlate independently of phylogeny for all amphibia (*12*), which is consistent with the phylogenetic independence of the correlation between genome size and species richness (see Figure 2). Together, these results support the proposal that genome size has a negative impact on species richness that is related to habitat range and diversity.

*Genome size is not correlated with latitude*

We next examined the relationship between latitude and genome size. Although latitude does not correlate with genome size, a pattern between genome size and latitude is nonetheless apparent (Figure 4D). Six of the ten salamander families inhabit the same range of latitudes between thirty-one and thirty-eight degrees from the equator. Two, Rhyacotritonidae and Dicamptodontidae, inhabit more elevated latitudes in a narrow range between forty and fort-six degrees in the Pacific Northwest.

The Ambystomatidae and Plethodontidae, in contrast, inhabit relatively lower latitudes ranging from twenty-nine to twenty-two degrees (Figure 4D), indicating a distribution skewed more toward Southwest North America and the tropics. These values are averages; and they therefore correspond to a wide range of latitudes comprising diverse biomes and ecosystems. With the exception of the Amphiumidae, the four outliers nevertheless have similar stem ages, and are younger than the six mid-latitude families (stem age: 125 Mya versus 165 Mya). The lower latitudes occupied by the Plethodontidae correspond to the adaptive radiation of the Bolitoglossinae situated in the tropics (*36*).

Chryptobranchidae and Sirenidae represent the median genome size, and are located in Figure 4D at thirty-one and thirty-five degrees latitude, respectively. In this range of latitudes average genome size varies the most: between 20 and 80 pg. In contrast, genome size at higher latitudes varies from 60 to 70 pg between the Rhyacotritonidae and the Dicamptodontidae, while it varies by about 15 % (30 to 35 pg) between the Ambystomatidae and the Plethodontidae at the lower latitudes. The large variation in genome size at middle latitudes most likely reflects the temperate origin of the ancestor of the ten extent salamander families (*13, 36*), which is consistent with their older stem ages compared to the four outlying families. Hence, salamander families of more recent origin occupy outlying latitudes North and South of the older families.

Although the Ambysomatidae, Plethodontidae, Rhyacotridontidae and Dicoamptodontidae all have similar stem ages, they differ significantly in terms of species richness (*27*). Consistent with a latitudinal gradient, the latter two are significantly less speciose than the Ambystomatidae and Plethodontidae. Average genome size, likewise, is nearly 2X larger than Plethodontidae or Ambystomatidae. Conversely, the Hynobidae and Salamandridae occupy similar latitudes in temporate zones; but have similar levels of species richness, suggesting that genome size is more closely associated with levels of species richness than either latitude or stem age.

*Ecological correlates with genome size: niche rate and temperature*

With the exception of latitude, the ecological factors examined above are all related and interact with each other in a phylogenetically independent manner. If the process of competitive exclusion operates geographically as well as ecologically (*37*), then the area that a family occupies will naturally increase as species richness increases. Similarly, larger areas are expected to have higher habitat diversity, and consequently clades with higher species richness should reflect higher niche rates (*38*). Niche rate refers to climatic changes in a given area over time, and is therefore believed to be related to niche breadth and heterogeneity (variance in niche-width). There should be then a definite relationship between the parameters of time, area and habitat diversity that acts to promote increases in species richness in a genome size dependent manner.

Consistent with salamander families of smaller C-values and higher species diversity occupying larger geographical areas, genome size is negatively correlated with climatic niche rate. Figure 5A shows a more uniform distribution between genome size and niche rate with the exception of the Ambystomatide, which have niche rates of nearly 2X that of families with similar genome sizes (< 50 pg). This indicates that other lineage-specific effects, for example narrower niches in the tropics (*39*), impact adaptive radiations. Despite these and other important lineage dependent factors, genome size nevertheless appears to influence niche rate in salamanders, suggesting that species with smaller genomes are more adaptable to climatic changes and fluctuations.

Interestingly, the Plethodontidae are the only family that exhibits a correlation of genome size with niche composition (abiotic factors of temperature and humidity). Using Pyron and Wien's data from their phylogenetically corrected principal component analysis (PCA) reveals that genome size in Plethodontidae (*12*), and only in that family, is negatively correlated with PC1 (see Figure 5B and Supplementary Figure 1), which is closely related to latitude and strongly dependent on temperature variables such as average annual temperature, temperature seasonality (negatively correlated with PC1) and temperature annual range (positively correlated with PC1). Moreover, PC1 shows a significant relationship with speciation and extinction rates. Within defined lineages, however, genome size does not appear to be correlated with PC1, perhaps because species within lineages such as the Bolitoglossinae and the Desmognathinae have similar C-values and inhabit, respectively, similar climatic niches.

Why genome size in this family increases as annual average temperatures increase and temperature ranges decrease is unclear. We note, however, that the Bolitoglossinae are the salamander lineages that have most successfully invaded the tropics, and are the most species rich tribe of Plethodontidae. They also have among the largest average genome size compared to other Plethodontidae and the slowest temperature seasonality niche rates and the narrowest niche breadths (*39*), which is believed to promote species richness via niche divergence (*40*). We conclude that in the Plethodontidae larger genome sizes and higher species richness correspond to more stable and higher average seasonal temperatures.

**Discussion**

*Does propensity to speciate vary with genome size?*

The correlations between crown age and species richness suggest that time to speciation in salamanders accounts for the observed species diversity in each family: older clades consistently have more species, and began radiating earlier relative to stem age than do less speciose clades (Figure 1; Figure 5C). This observation is consistent with the finding that crown age rather than stem age explains species richness across the eukaryotic tree of life (*41*). Here, we have extended the observation on time-to-speciation and species richness to include family level variations in genome size: more speciose families, which correspond to adaptive radiations that began earlier with respect to the stem age, have consistently smaller average genome sizes in salamanders (Figure 1; Figure 5D). No correlation, in contrast, exists between stem age and C-value (Figure 4A). A possible explanation for the correlation between stem-to-crown age and C-value might be that the most recent common ancestor (MRCA) of species in families with smaller genomes had a higher propensity to speciate than the MRCA of species in families with larger genomes, and consequently the clade began diverging earlier than its sister taxon.

*Two alternative modes of genome size evolution and species diversity?*

Two possible interpretations, among others, might explain the observation that salamander families with shorter stem-to-crown age intervals tend to be more speciose (Figure 5C). First, these families have been radiating longer relative to stem age, and changes in genome size have either enhanced or diminished speciation events. The Hynobidae and Cryptobranchidae, for example, have the same stem age, being sister taxa; but the Cryptobranchidae have younger crown ages, which suggests the Cryptobranchidae experienced a long period of evolutionary stasis before they began diverging (Figure 1). Consistent with this proposal, morphological traits in salamanders are evolving at a rate that is correlated with species richness: families experiencing slower evolving phenotypic traits are correspondingly less speciose (*42*), which might imply a long period of evolutionary stasis before adaptive radiation began.

The Hynobidae, likewise, might have experienced a relatively shorter period of evolutionary stasis before they began radiating. The Hynobidae, for example, have been diverging for some 60 million years longer than the Cryptobranchidae (crown age: 134.7 vs. 67 Mya, respectively). The earlier Hynobidae radiation might therefore have been associated with genome size fluctuations that resulted in an overall reduction in genome size, which could explain the smaller average C-value in this family compared to its sister taxon. Conversely, the apparent long period of stasis the Cryptobranchidae experienced might have been associated with a corresponding increase in genome size.

Deciding between these two scenarios (genome reduction or genome expansion) depends on the MRCA's genome size at the time the two lineages diverged; but whether the MRCA had a larger or smaller genome relative to the sister taxa leads to the same conclusion: reductions in genome size either promote species richness by enhancing rates of speciation (or by slowing extinction rates), or amplifications in genome size diminish species richness by slowing speciation rates (or by increasing extinction rates). Together, a longer time to speciation and lower species richness in Cryptobranchidae compared to the Hynobidae implies that genome size affects rates of evolution either directly or indirectly in salamanders.

The second, related interpretation concerns rates of molecular evolution and diversification. We found a weaker negative correlation between species diversification rates and C-value (not shown); diversification rates, however, reflect both speciation and extinction rates. Which of these two factors principally influences current species richness is difficult to assess. Nevertheless, extant species diversity is likely to decrease if the family is not evolving fast enough (with sufficiently high speciation rates) to compensate for the probability of extinction, which increases over time. Consequently, only those families/clades that can generate viable levels of genetic and phenotypic variation will survive the probability of extinction long enough to form extant species. Such families, presumably, will have comparatively higher rates of neutral mutation that depend on genome size.

According to this interpretation, the younger crown age in families with larger genomes reflects instead slower evolutionary rates in those families rather than more recent adaptive radiations: either the emergence of distinct species took longer in families with larger genomes, or the emergence of distinct species occurred earlier and faster in families with smaller genomes. This interpretation is consistent with a mechanistic interaction between genome size, molecular rates of evolution and extant species richness. We suggest that the time-dependent (molecular clock) variations in genome size underlie, mechanistically, the molecular origin of species in salamanders (see below). This suggestion finds tentative support in the close association found here between the standardized contrasts in genome size and standardized contrasts in species richness (Figure 2B).

*Does variance in species richness in urodela reflect variance in mutation rates?*

The Plethodontidae represent a potential exception to these two alternative genome size dependent interpretations. Consistent with the above proposal, the Plethodontidae have the highest species richness and the largest range in genome size (10 pg to over 70 pg). Tribes with larger genomes such as the Bolitoglossinae and Plethodontinae, however, are more speciose than other tribes with smaller genomes (*27*). Thus, at the genus level in Plethodontidae no clear correlations, either positive or negative, exist between genome size and species richness (not shown). Interestingly, among the Plethodontidae, family niche width explains variations in diversification rates more substantially than does species niche width (*40*), indicating that these diverse effects on species richness (genome size and niche width) are apparent primarily at the family rather than the genus level.

The observation of larger variances in genome size and species richness in Plethodontidae is consistent with our earlier reported finding of a larger variance in mutation rates in the *rag1* gene within the Plethodontidae compared to the other salamander families (*43*). Hence, the Plethodontidae are the most species rich family and have the highest variance in both genome sizes and mutation rates compared to the other salamander families. Indeed, the previous study on mutation rates showed that the Bolitoglossinae have some of the fastest rates of evolution at the molecular level (*rag1*); while the Desmognathinae, with some of the smallest genome sizes, have rates of molecular evolution that are substantially less than tribes with larger C-values. These observations on the variance in rates of molecular evolution and genome size in salamanders remain, however, to be confirmed by more extensive studies concerning the impact of genome size on rates of molecular evolution.

These respective findings nevertheless suggest that rates of molecular evolution and speciation are correlated and could even influence each other. What might be the relationship between mutation rates, rates of speciation and species richness? Changes in genome size reflect *de facto* mutations that are the consequences of DNA repair mechanisms, transposon proliferation and other forms of genomic instability (*3*). The ancestral salamander genome was comparatively small -- approximately 3 pg -- which is similar to the average mammalian genome size (*44*). Salamanders have therefore experienced massive genome amplification during the course of evolution. Comparing Hynobidae and Cryptobranchidae, for example, suggests that these closely related families experienced significantly different mutation rates as they evolved, and that these different rates influence the different levels of species richness in these two families. The relationship between mutation rate and speciation rate remains, however, a point of considerable controversy (*45-48*).

*Junk DNA: driving reproductive isolation and speciation?*

Vertebrate species such as urodela differ more in their respective amounts of non-coding DNA than in their respective number of genes: the C-value enigma (*49, 50*). The species specific differences in non-coding DNA therefore distinguishes species perhaps as much as differences in their coding

DNA. What role then could junk DNA play in distinguishing between different species and higher taxon levels? We propose, in accordance with other authors (*51, 52*), that the wide variation in non-coding DNA in eukaryotic genomes implies that non-coding DNA plays a potential role in the mechanisms driving speciation/reproductive isolation and in influencing context dependent rates of mutation and species divergence. Heterochromatin, for example, which is composed of neutrally evolving sequences, is itself functional and actively maintained in the genome, and appears to play a role in speciation (*52*).

Junk DNA has been previously proposed to represent a second molecular clock (*53*). Accordingly, the rate of change in genome size and organization (synteny and karyotype) in diverging populations might set the rate at which populations become reproductively isolated (see Figure 2 and 4B), resulting in the distinct populations on which natural selection can subsequently act. A similar proposal has recently been made suggesting that the split between two lineages in a phylogenetic tree corresponds to the moment of reproductive isolation rather than speciation. A long period of time follows the split allowing enough genetic incompatibility to accumulate randomly between the two populations before adaptive changes result in phenotypically differentiated sister species (*41*). Random changes in the levels and organization of junk DNA therefore appear to play a role in generating the genic and genomic incompatibilities leading to the reproductive isolation the entails speciation. In that regard, the distinction made between neutral and adaptive mutations might be more of heuristic rather than explanatory value in interpreting mutation rates and their impact on genome evolution, reproductive isolation and the origin of species.

Evidence is emerging in support of such a view. Genetic drift, via the insertion and deletion of neutral DNA, modifies the chromatin and mutational context of genes both across species and during development, and provides the raw material, molecularly and physiologically, on which natural selection can operate for improving overall organism and species fitness (*54, 55*). A recent hypothesis proposes that variations in intra-genomic mutation rates, which are mediated by DNA replication timing and chromatin context (heterochromatin versus euchromatin), can explain gene positions according to gene ontology in a manner that reflects the kinetics of differential DNA repair mechanisms (non-homologous end-joining versus homologous recombination) (*56*). More recently a study, which confirms and extends earlier studies (*57, 58*), has demonstrated that gene and DNA sequence age correlate with replication timing: DNA copy number variants (CNV) of older origin replicate early, whereas more recently acquired CNVs replicate late (*59*). Hence, genes associated with late replicating CNVs are expected to experience correspondingly higher rates of molecular evolution and adaptation in a manner dependent on DNA replication errors (*60*).

*Conclusion*

Our principal conclusions are threefold: 1) genome size in salamanders correlates with a number of important ecological variables such as niche rate and area; 2) genome size influences species richness at the family level; and 3) standardized contrasts in genome size are strongly associated with standardized contrasts in species richness; if standardized contrasts can be interpreted as a rate in evolutionary terms (*5*), then the rate at which genome size varies in evolutionary time (branch lengths) correlates closely with the rate at which species richness varies, suggesting a significant interaction between variations in genome size and evolutionary fates in salamander families. Together, our findings indicate that genome size is responsive to a number of ecological factors, and in conjunction with these factors contributes to explaining species richness in the respective families.

Additionally, these findings indicate that while genome size evolves neutrally, the impact of genome size on species richness is non-random and constitutes a component in the array of phenotypes on which natural selection acts. Non-adaptive DNA sequences might therefore evolve neutrally in the short term while nonetheless entailing non-neutral sequence-independent effects in the long term. The proposals advanced here are based more on argument at this point than evidence, but they can soon be put to the test after more eukaryotic genomes have been sequenced (*61*). The mystery of non-coding, junk DNA and what exactly it is doing in the genome will likely persist, however, for some time to come.

**Materials and methods**

Genome sizes were obtained from the Animal Genome Size Database (*1*). The data on crown age, stem age, area, niche rate, latitude and PC1 were obtained from Pyron and Wiens (*12*). Independent contrasts were carried out in R using the ape library (http://www.r-phylo.org/wiki/HowTo/Phylogenetic_Independent_Contrasts) based on the branch lengths of the tree shown in Figure 1 as obtained from the tree of Pyron and Wiens (*27*). The regression of the independent contrasts was forced through the origin. Multiple regression analysis was carried out in R and Excel.

**Acknowledgements**

BS is supported by a grant from Human Frontier Science Program (RGY0079). JH benefited from support from John Bechhoefer's lab, Physics Department, Simon Fraser University.

**Table 1**

**Linear correlation**

|  | Residual Sum of Squares | Adj. R-Squared | Intercept | Standard Error | Slope | Standard Error | F Value | Prob>F |
|---|---|---|---|---|---|---|---|---|
| Log(STV) vs ln(species) | 0.70 | 0.44 | 1.9 | 0.2 | -0.16 | 0.06 | 7.99 | 0.02 |
| Log(C value) vs ln(species) | 0.12 | 0.58 | 1.90 | 0.07 | -0.09 | 0.02 | 13.52 | 0.006 |
| Log(C value) vs log(STV) | 0.24 | 0.19 | 1.3 | 0.2 | 0.3 | 0.1 | 3.08 | 0.1 |

*Independent Contrasts*

|  | Residual Sum of Squares | Adj. R-Squared | Intercept | Standard Error | Slope | Standard Error | F Value | Prob>F |
|---|---|---|---|---|---|---|---|---|
| *Log(STV) vs ln(species)* | 2.97 | 0.52 | 0.15 | 0.23 | -0.16 | 0.05 | 9.61 | 0.02 |
| *Log( Cvalue) vs ln(species)* | 0.54 | 0.84 | -0.14 | 0.07 | -0.11 | 0.02 | 43 | 0.0003 |
| *Log(C value) vs log(STV)* | 1.06 | 0.51 | -0.19 | 0.14 | 0.45 | 0.15 | 9.49 | 0.02 |

**Table 2**

*Multiple regression analysis (species richness)*

*Residual standard error: 0.99*
*Adj. $R^2$: 0.67*
*F Value: 10*
*p-value: 0.008*

|  | Estimate | Std. Error | t value | Pr(>|t|) |
|---|---|---|---|---|
| *(Intercept)* | 14 | 3 | 4.8 | 0.002 |
| *Log(STV)* | -1.7 | 0.9 | 1.8 | 0.1 |
| *Log(C value)* | -5 | 2 | -2.5 | 0.03 |

**Table 3**

*Partial correlation analysis (species richness)*

| Predictor | $R^2$ | β weights | Unique | Common | Total | Unique % Total | Common % Total |
|---|---|---|---|---|---|---|---|
|  | 0.69 |  |  |  |  |  |  |
| *Log(STV)* |  | -0.46 | 0.11 | 0.38 | 0.49 | 16 | 52 |
| *Log(C value)* |  | -0.58 | 0.24 | 0.38 | 0.62 | 33 | 52 |

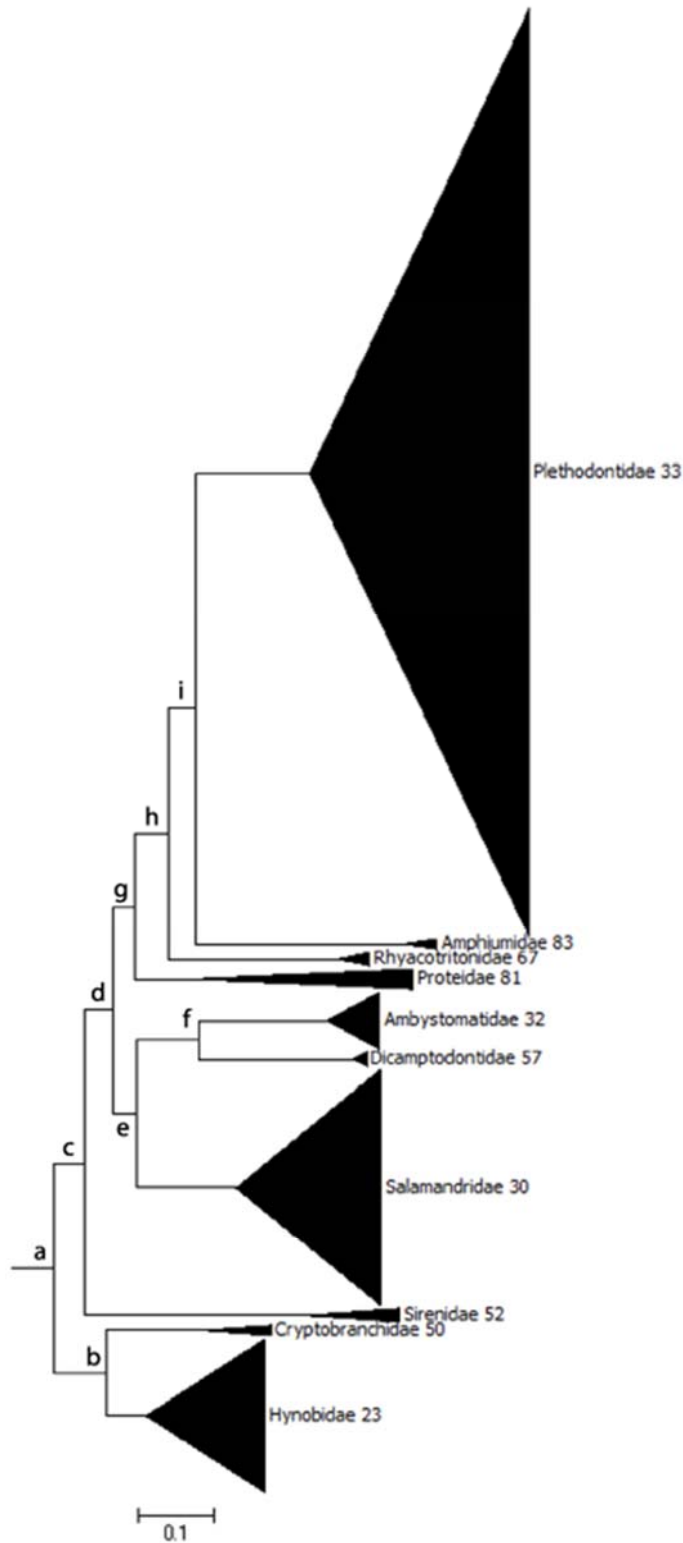

**Figure 1.** Phylogenetic tree of the ten urodela families obtained from Pyron and Wiens (*27*). The average C-value is shown next to the family name. The size of the triangle (black) is proportional to the number of species subtending the crown of each clade. The branch length is indicated by the scale bar. The letters denote the node identities used in the analysis of phylogenetically independent contrasts (Figures 2 and 3). Six families form three sister-pair taxa: nodes i, f and b, which correspond to the most recent common ancestor (MRCA) of the subtending families.

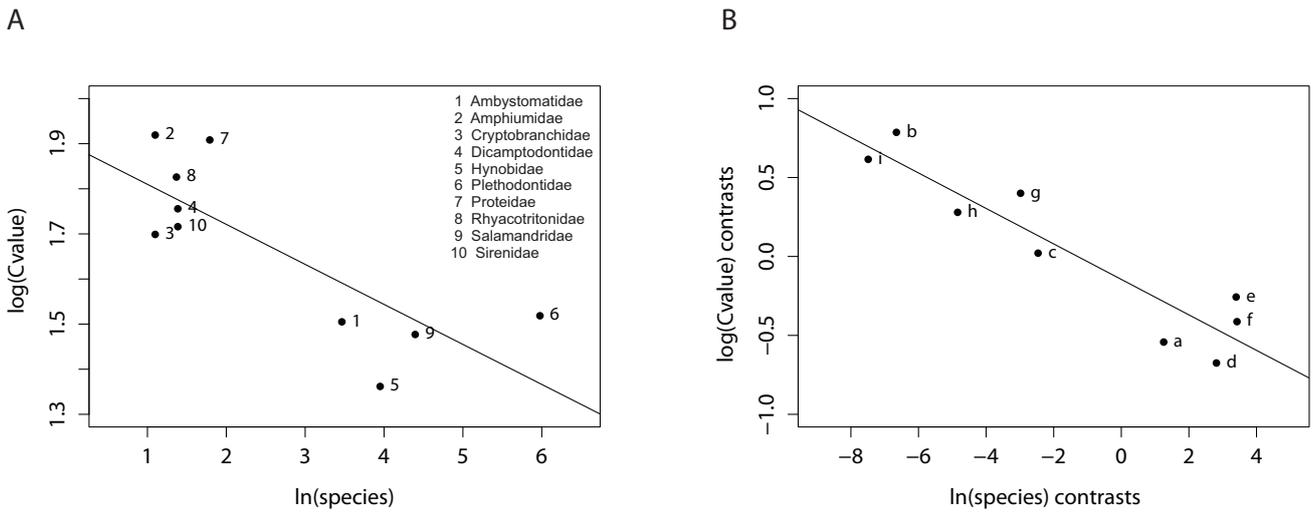

**Figure 2.** Species richness and C-value are negatively correlated at the family level in salamanders. A) Regression between ln(species) (species richness) and log(C-value) (genome size) revealing two distinct classes of salamander (> 50 pg and < 50 pg). Adjusted $R^2$: 0.58; P-value: 0.006; Slope: -0.09 ± 0.02 B) Contrasts scaled using the expected variances: adjusted $R^2$: 0.84; P-value: 0.003; Slope: -0.11 ± 0.02. Numbers correspond to the different families and letters to the nodes of the tree in Figure 1.

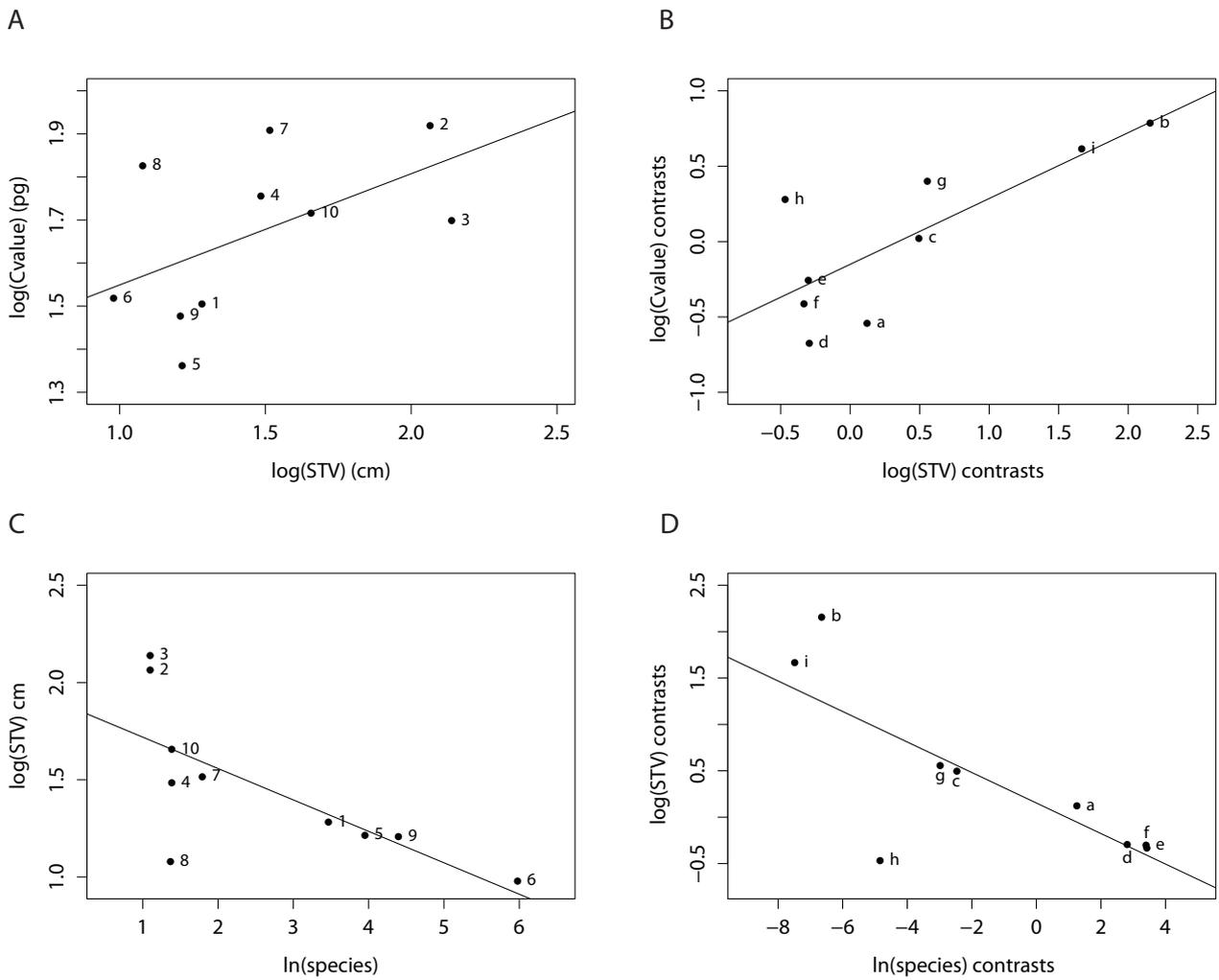

**Figure 3.** Genome size and body size are positively correlated at the family level in salamanders, while species richness and body size are negatively correlated. A) log(STV) (snout to vent length) vs log(C value) (genome size). Adjusted $R^2$: 0.19; P-value: 0.1; Slope: 0.26 ± 0.15. B) Contrasts scaled using the expected variances: adjusted $R^2$: 0.55; P-value: 0.01; Slope: 0.4 ± 0.1. C) log(STV) (snout to vent length) vs ln(species) (species richness). Adjusted $R^2$: 0.44; P-value: 0.02; Slope: -0.16 ± 0.06. B) Contrasts scaled using the expected variances: adjusted $R^2$: 0.52; P-value: 0.02; Slope: -0.16 ± 0.05. Numbers correspond to the different families as shown in Figure 2 and letters to the nodes of the tree in Figure 1.

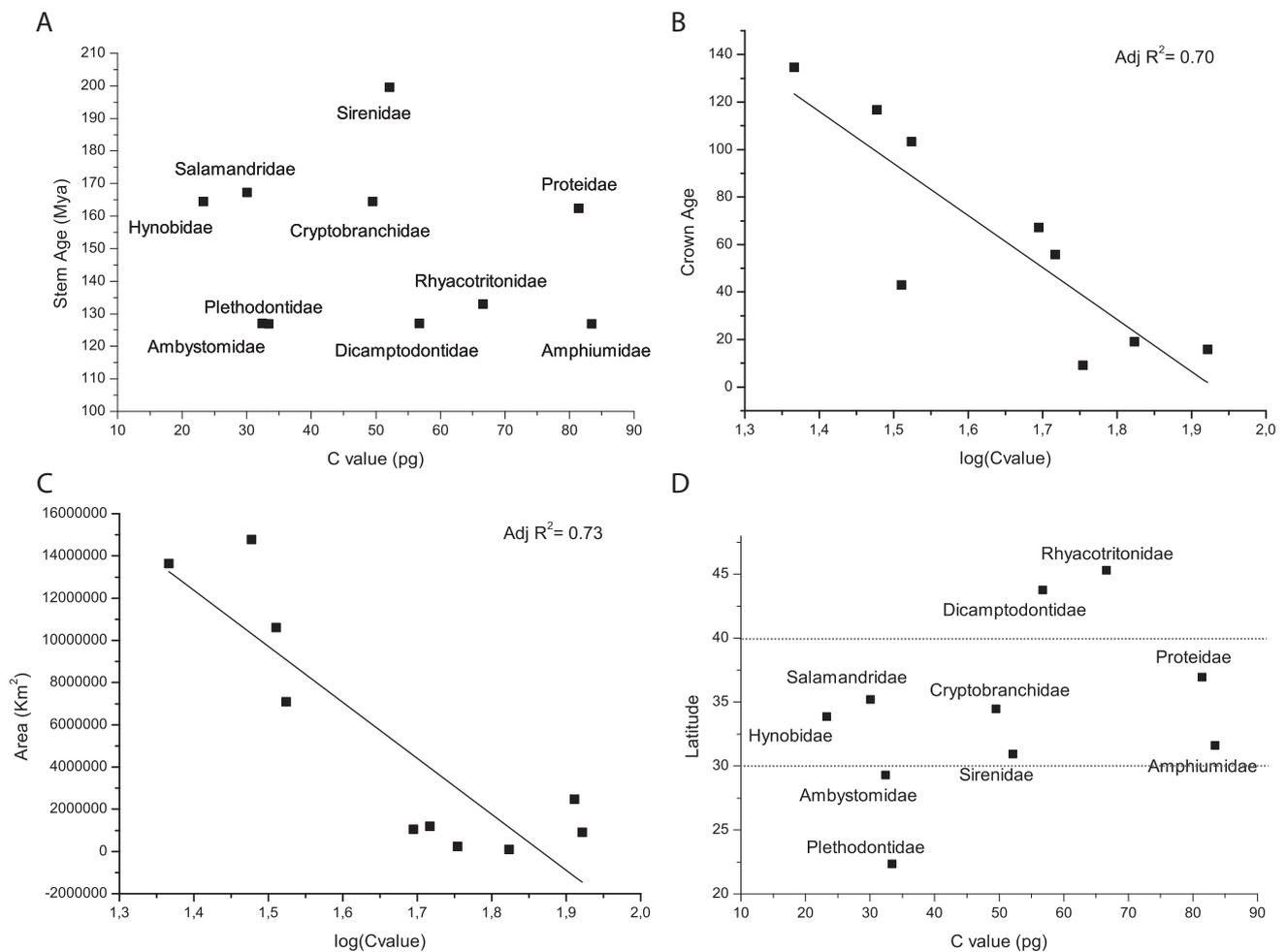

**Figure 4.** Correlations between C-value and time and area. A) Regression of stem age on C-value reveals no correlation between family age and genome size for the ten named families. B) Regression of crown age (time) on log(C-value) (genome size) reveals a strong correlation between clade age and C-value ($R^2 = 0.70$). C) Regression of area on log(C-value) (genome size) reveals a similarly strong correlation between the geographic ranges occupied by the ten families and their respective C-values ($R^2 = 0.73$). D) Regression of latitude on C-value reveals no correlation between family genome size and geographic position. Note, however, that the species rich Ambystomatide and Plethodontidae occupy lower latitudinal positions (22 to 28° N) while the Dicamptodontidae and Rhyacotridontidae, which have larger average C-values and lower species richness, occupy higher latitudinal positions (42 to 46° N). Most families occupy mid-latitude positions (32 to 36° N), reflecting the temperate origin of urodela.

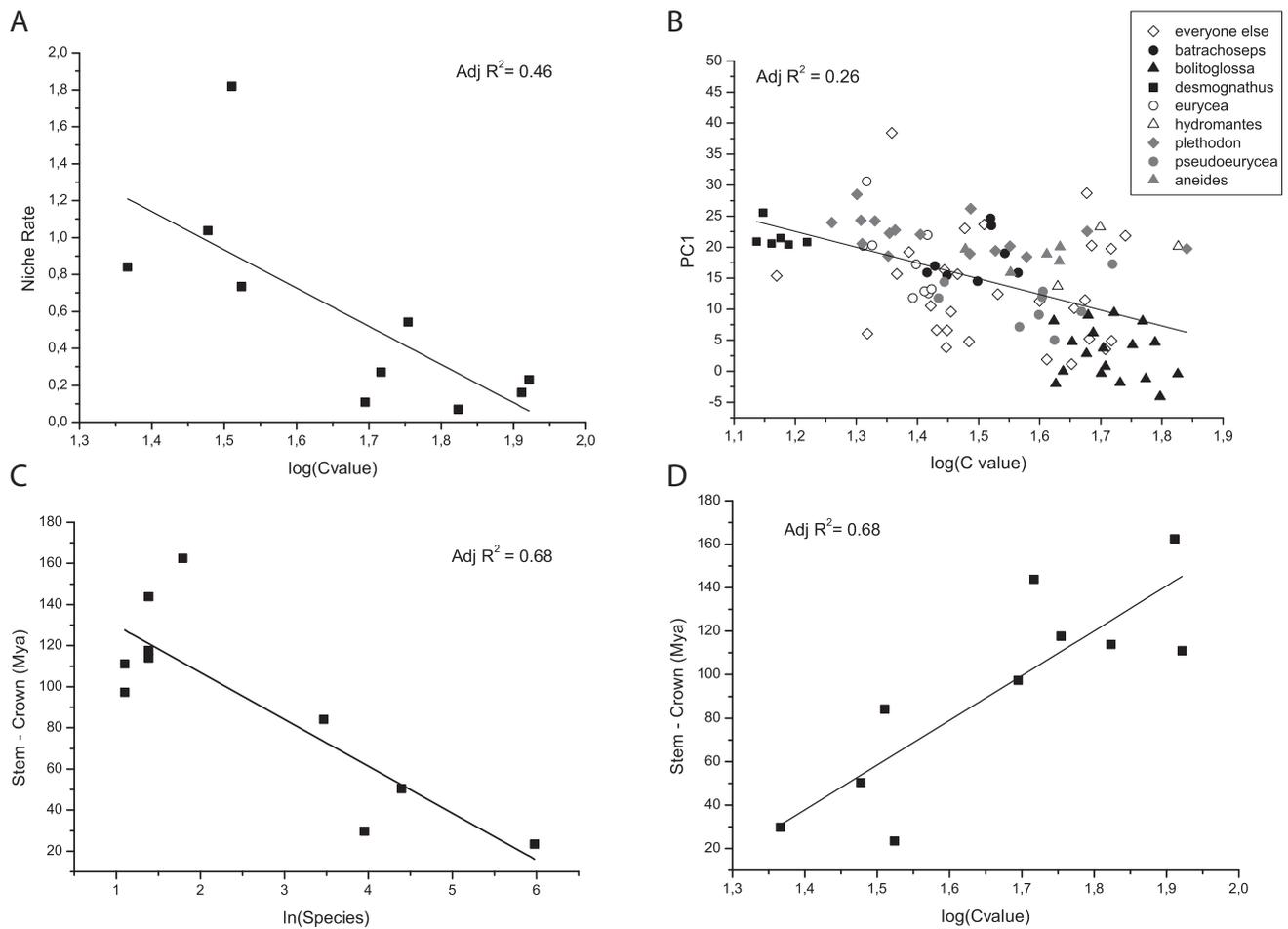

**Figure 5.** Niche rate and evolutionary history (time between stem age and crown age). A) Regression of niche rate on log(C-value) (genome size) reveals a strong negative correlation between rates of climactic niche evolution and differences in genome size ($R^2 = 0.48$). B) Regression of PC1 (temperature variables influencing climactic niche) on log(C-value) (genome size) in Plethodontidae ($R^2 = 0.26$). C) Regression of temporal difference between stem age and crown age on species richness reveals that families experiencing longer periods of evolutionary and adaptive inertia are less species rich ($R^2 = 0.68$). D) Regression of stem age vs. crown age on log(C-value) (genome size) reveals that families experiencing higher levels of evolutionary inertia tend to have larger genomes ($R^2 = 0.68$). Figures 4C and 4D indicate that species richness and C-value are tightly coupled and negatively associated in urodela.

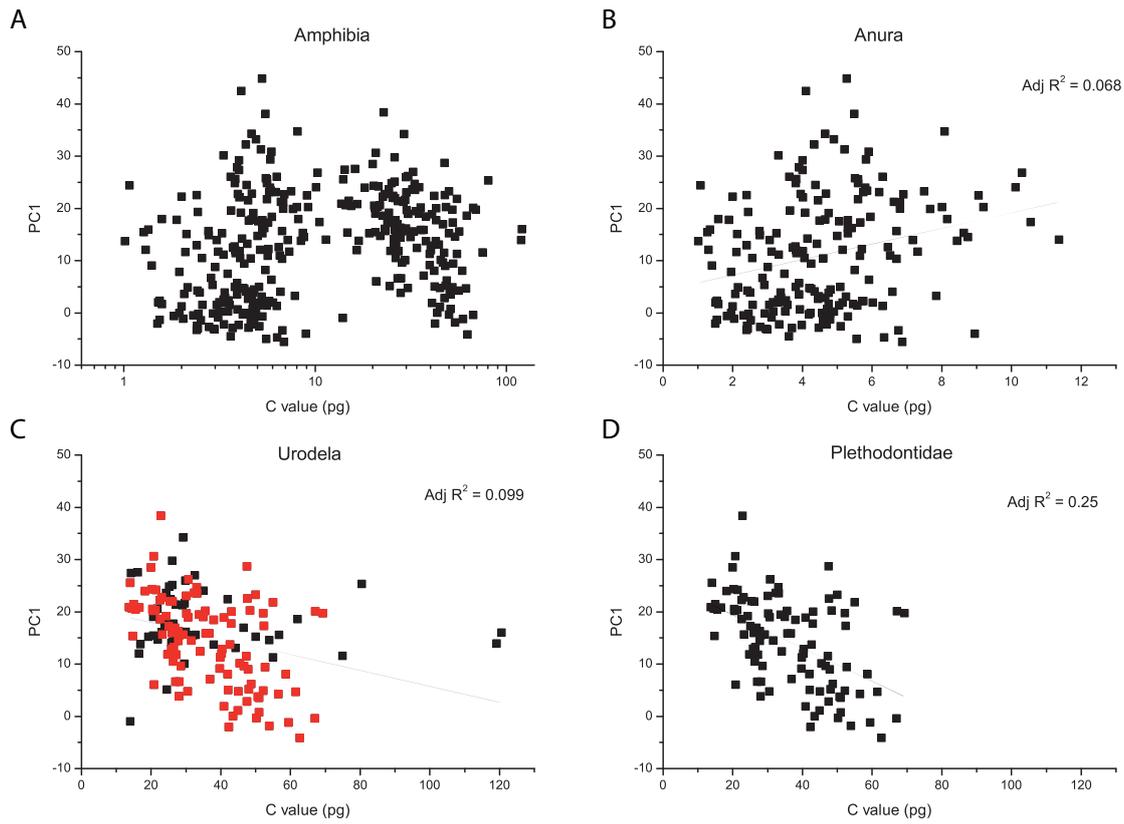

**Supplementary Figure 1.** PC1 as a function of genome size in Amphibia (A), within Anura (B), Urodela (C), where the plethodontidae are shown in red and Plethodontidae alone (D).